\documentclass[twocolumn,showpacs,preprintnumbers,amsmath,amssymb]{revtex4}

\usepackage{graphicx}
\usepackage{enumerate}
\usepackage{amsmath}
\usepackage{bm}

\newcommand{\bra}[1]{\langle #1 |}
\newcommand{\ket}[1]{| #1 \rangle}

\begin{document}

\title{A scalable cavity-QED-based scheme of generating entanglement\\of atoms and of cavity fields}

\author{Jaehak Lee$^1$, Jiyong Park$^1$, Sang Min Lee$^1$, Hai-Woong Lee$^1$, and Ashfaq H. Khosa$^2$}
\affiliation{$^{1}$Department of Physics, Korea Advanced Institute of Science and Technology, Daejeon 305-701, Korea\\
$^{2}$Centre for Quantum Physics, COMSATS Institute of Information Technology, Islamabad, Pakistan}

\begin{abstract}
We propose a cavity-QED-based scheme of generating entanglement between atoms. The scheme is scalable to an arbitrary number of atoms, and can be used to generate a variety of multipartite entangled states such as the GHZ, {\it W} and cluster states. Furthermore, with a role switching of atoms with photons, the scheme can be used to generate entanglement between cavity fields. We also introduce a scheme that can generate an arbitrary multipartite field graph state.
\end{abstract}


\maketitle

\section{INTRODUCTION}

Quantum entanglement plays a central role in many applications of quantum-information science such as quantum teleportation \cite{bib:tele}, quantum cryptography \cite{bib:crypt}, and quantum computation \cite{bib:compu}. Many schemes of generating various types of entanglement have thus been proposed \cite{bib:prev1, bib:prev2, bib:prev3, bib:prev4, bib:prev5, bib:prev6} and some experimentally demonstrated in the past \cite{bib:exp1, bib:exp2, bib:exp3, bib:exp4, bib:exp5, bib:exp6}. The most commonly used source of entanglement is spontaneous parametric down conversion, which generates polarization entanglement between two photons. It is, however, not as straightforward to generate entanglement in massive particles as in photons. With recent advances in cavity QED technologies, cavity-QED-based schemes are an attractive candidate for generation of entanglement between atoms or ions. The first successful generation of entanglement between two atoms has indeed been achieved in a cavity QED experiment \cite{bib:exp1}.

With many impressive progresses witnessed in recent years \cite{bib:exp1, bib:exp2, bib:exp3}, it no longer seems to represent a high technological challenge to generate bipartite entanglement in atoms as well as in photons. Generation of multipartite entanglement, however, remains to be a difficult task, despite some notable recent achievements \cite{bib:exp4, bib:exp5} along this direction. Evidently, multipartite entanglement is an essential ingredient for quantum-information processing in a network environment, for example, quantum communication between the communication center and multi-users. Multipartite entanglement is also of utmost importance in one-way quantum computing \cite{bib:compu}, where quantum computation is achieved first by preparing qubits in the cluster state, a particular class of multipartite entangled state, and then by performing single qubit measurements. It therefore is still a desirable task to develop an experimentally feasible method, which generates different types of multipartite entanglement in atoms \cite{bib:exp5} or in photons \cite{bib:exp4}.

In this paper we propose a cavity-QED-based scheme that generates entanglement between atoms. The scheme makes use of linear optical devices and has an ideal success probability of 100\%. The scheme is scalable, i.e., it can generate multipartite entanglement among an arbitrary number of atoms. A particular merit of the scheme is its versatility, as it can be tailored to generate different types of multipartite entangled states such as the GHZ state \cite{bib:GHZ}, W state \cite{bib:W} and the cluster state \cite{bib:compu} with slight rearrangements of the configuration. Furthermore, with switching of the role played by atoms and photons, the scheme can be used to generate entanglement between cavity fields \cite{bib:prev6, bib:exp6} instead of between atoms.

\section{BASIC IDEA}

\begin{figure}[b]
\centering \includegraphics[height=0.8\columnwidth, angle=270]{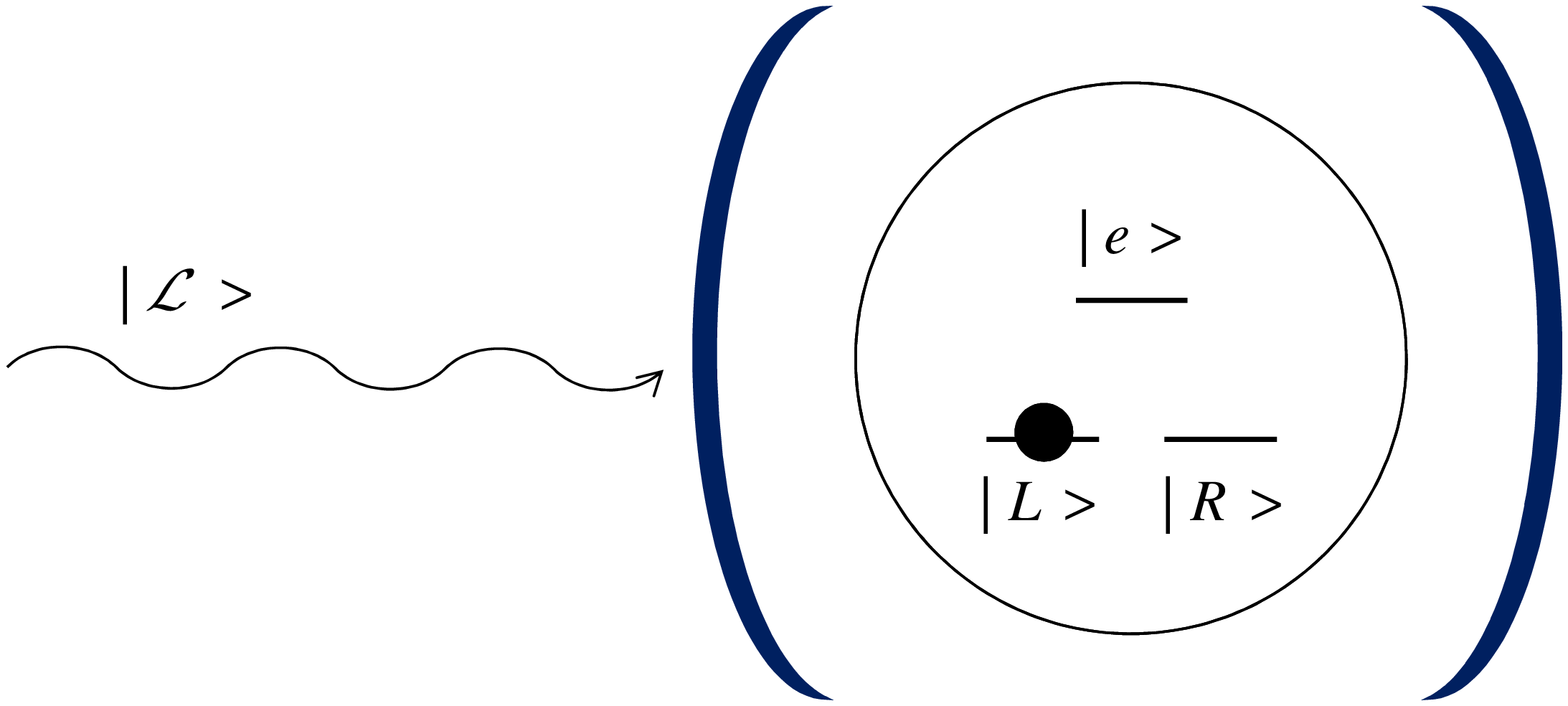}
\caption{\label{fig:atomblock} The basic building block of the proposed scheme. The adiabatic interaction between a three-level atom prepared in $\ket{L}$ in a cavity and a left-circularly polarized photon $\ket{{\cal L}}$ transforms the system to the atom in $\ket{R}$ and a right-circularly polarized photon $\ket{{\cal R}}$ and vice versa.}
\end{figure}

The basic building block of our scheme is an atom-photon system of Figure \ref{fig:atomblock}, which performs a NOT operation on an atom trapped in a cavity.
The atom is a $\Lambda$-type three-level atom with two lower states $\ket{L}$ and $\ket{R}$ as the basic qubit states. The states $\ket{e}$ and $\ket{L}$($\ket{R}$) are coupled via a left- (right-) circularly polarized photon $\ket{{\cal L}}$($\ket{{\cal R}}$). A standard analysis of the cavity input-output process \cite{bib:io}, which is presented below, yields that, in the adiabatic limit, the state transformation $\ket{L}\ket{{\cal L}}\leftrightarrow\ket{R}\ket{{\cal R}}$ is accomplished with an ideal success probability of $\sim$100\% \cite{bib:prev4}. This provides a simple way of flipping atomic qubits $\ket{L}\leftrightarrow\ket{R}$ through interaction with a photon.

Let us assume that the atom trapped inside a cavity in vacuum is prepared initially in state $\ket{L}$ and a left-circularly polarized photon enters the cavity through a fiber coupled to the cavity. The initial state of the system, atom+cavity+fiber, can be written as $ \ket{\psi(0)} = \ket{L;0,0}\ket{\phi_{\cal L}(0)} $, where the symbols in the first ket on the right side denote the state of the atom and the number of ${\cal L}$ photons and ${\cal R}$ photons, respectively, and the second ket is used to indicate the state of photons in the fiber. The state at any later time $ t>0 $ can be written as
\begin{equation} \label{eq:state} \begin{split} & \ket{\psi(t)} \\
& = c_L(t)\ket{L;1,0}\ket{0}+c_R(t)\ket{R;0,1}\ket{0}+c_e(t)\ket{e;0,0}\ket{0} \\
& +c_{L,\textrm{out}}(t)\ket{L;0,0}\ket{\phi_{\cal L}(t)}+c_{R,\textrm{out}}(t)\ket{R;0,0}\ket{\phi_{\cal R}(t)}
\end{split} \end{equation}
The interaction Hamiltonian governing the time evolution of the system is given by
\begin{equation} \label{eq:H} \begin{split}
H = & i\hbar\sum_{\lambda=L,R}\left(g_{\lambda}a_{\lambda}\sigma_{\lambda}^{+}-g_{\lambda}a_{\lambda}^{\dagger}\sigma_{\lambda}^{-}\right) \\
& + i\hbar\sum_{\lambda=L,R}\int_{-\infty}^{\infty}d\omega\sqrt{\frac{\kappa}{2\pi}}\left[b_{\lambda}^{\dagger}(\omega)a_{\lambda}-b_{\lambda}(\omega)a_{\lambda}^{\dagger}\right]
\end{split} \end{equation}
where $ \sigma_{\lambda}^{+} = \ket{e}\bra{\lambda} $, $ \sigma_{\lambda}^{-} = \ket{\lambda}\bra{e} $, $a_{\lambda}$ and $b_{\lambda}(\omega)$, respectively, are the annihilation operators for the photon of polarization $\lambda$ in the cavity and for the photon of polarization $\lambda$ and frequency $\omega$ in the fiber, $g_{\lambda}$ represents the atom-cavity field coupling constant, and $\kappa$ is the cavity decay rate. Clearly, the first term on the right side of Eq. (\ref{eq:H}) describes the atom-cavity field interaction and the second term the cavity-fiber coupling.

In order to describe the cavity input-output process, we make use of the input and output operators $b_{\lambda,\textrm{in}}(t)$ and $b_{\lambda,\textrm{out}}(t)$ \cite{bib:io} defined as
\begin{subequations} \label{eq:io} \begin{align}
b_{\lambda,\textrm{in}}(t) & = \frac{1}{\sqrt{2\pi}}\int d\omega b_{\lambda,0}(\omega)e^{-i\omega (t-t_0)} \\
b_{\lambda,\textrm{out}}(t) & = \frac{1}{\sqrt{2\pi}}\int d\omega b_{\lambda,1}(\omega)e^{-i\omega (t-t_1)}
\end{align} \end{subequations}
where $b_{\lambda,0}(\omega)$ and $b_{\lambda,1}(\omega)$ are the values of $b(\omega)$ at the initial time $ t=t_0(<t) $ and at the final time $ t=t_1(>t) $, respectively. The amplitude of the input and output pulses can then be expressed in terms of the input and output operators as
\begin{subequations} \label{eq:amp} \begin{align}
f_{\lambda,\textrm{in}}(t-t_0) & = \bra{\lambda;0,0}\bra{0}b_{\lambda,\textrm{in}}(t)\ket{\psi(0)} \\
f_{\lambda,\textrm{out}}(t_1-t) & = \bra{\lambda;0,0}\bra{0}b_{\lambda,\textrm{out}}(t)\ket{\psi(0)}
\end{align} \end{subequations}
The input and output amplitudes satisfy the identity
\begin{equation} \label{eq:bc} f_{\lambda,\textrm{out}}(t) = f_{\lambda,\textrm{in}}(t) + \sqrt{\kappa}c_{\lambda} \end{equation}
Through a straightforward algebra, we obtain a set of differential equations for the probability amplitudes, which read
\begin{subequations} \label{eq:de} \begin{align}
\dot{c_L} & = -\frac{\kappa}{2}c_L-g_Lc_e-\sqrt{\kappa}f_{{\cal L},\textrm{in}}(t) \\
\dot{c_R} & = -\frac{\kappa}{2}c_R-g_Rc_e \\
\dot{c_e} & = g_Lc_L+g_Rc_R
\end{align} \end{subequations}
Taking the adiabatic limit, i.e., setting the derivatives $\dot{c_L}$, $\dot{c_R}$ and $\dot{c_e}$ equal to zero, we obtain from Eqs. (\ref{eq:bc}) and (\ref{eq:de})
\begin{subequations} \label{eq:out} \begin{align}
f_{{\cal L},\textrm{out}}(t) & = \left(1-\frac{2g_R^2}{g_L^2+g_R^2}\right)f_{{\cal L},\textrm{in}}(t) \\
f_{{\cal R},\textrm{out}}(t) & = \frac{2g_Lg_R}{g_L^2+g_R^2}f_{{\cal L},\textrm{in}}(t)
\end{align} \end{subequations}
Eqs. (\ref{eq:out}) indicate that, when $g_L=g_R$, we have $ f_{{\cal L},\textrm{out}}(t) = 0 $ and $ f_{{\cal R},\textrm{out}}(t) = f_{{\cal L},\textrm{in}}(t) $. Thus, the transformation of the state from $\ket{L}\ket{{\cal L}}$ to $\ket{R}\ket{{\cal R}}$ is accomplished, i.e., the desired atomic state flip $\ket{L}\to\ket{R}$ is accomplished with 100\% success probability, $P=100\%$, in the adiabatic limit.

\begin{figure}
\centering \includegraphics[height=0.8\columnwidth, angle=270]{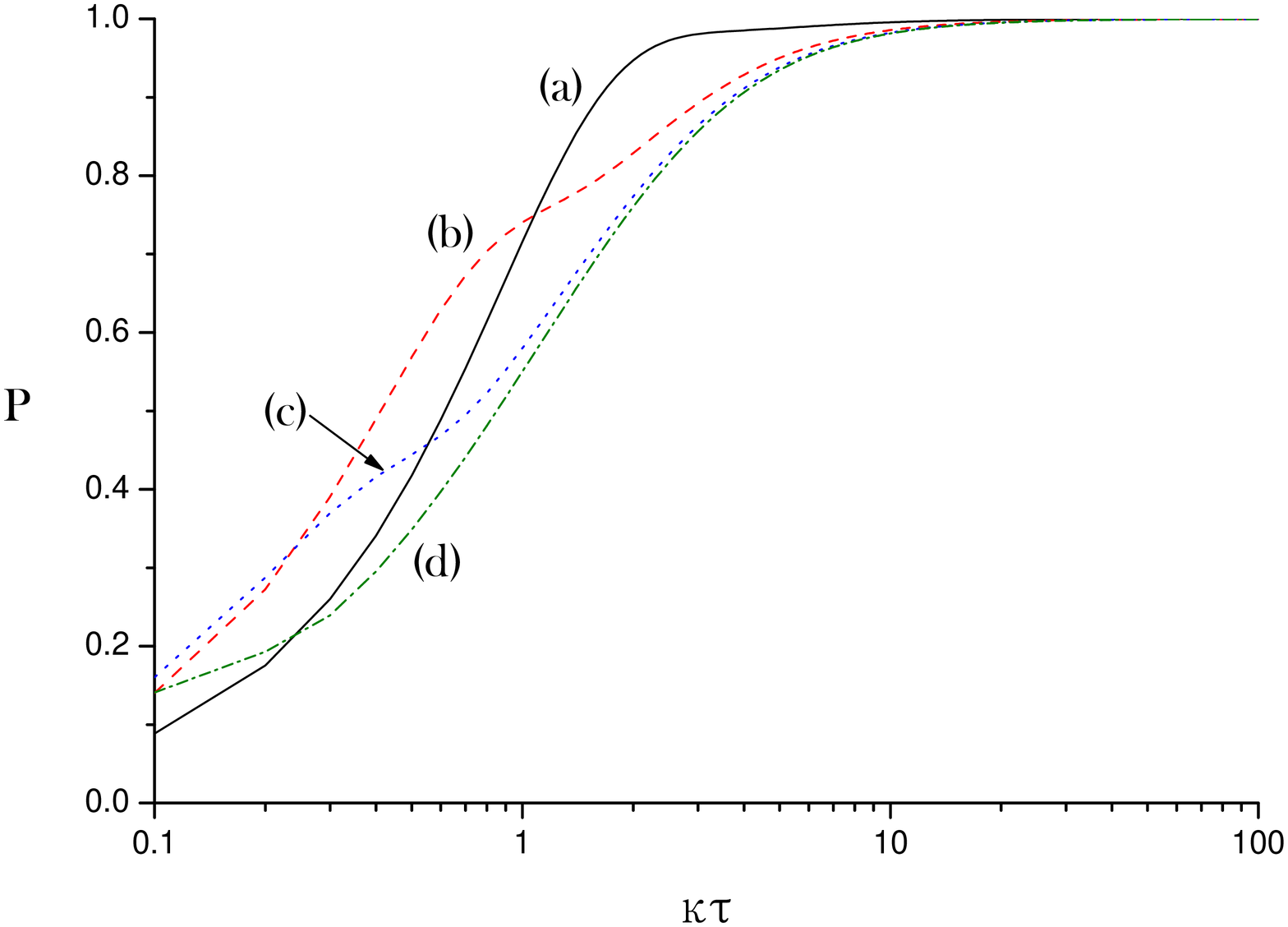}
\caption{\label{fig:adiabatic} The flip probability $P$ vs. the temporal width $\tau$ of a Gaussian pulse for the case $ g/\kappa = $ (a)0.5, (b)1.0, (c)2.0 and (d)5.0. It is assumed that $ g_L = g_R \equiv g $.}
\end{figure}

In order to assess the adiabaticity condition, we have computed the flip probability $P$ by numerically integrating Eqs. (\ref{eq:de}) for the case when the input pulse is of Gaussian shape, $ f_{L,\textrm{in}}(t) = \sqrt{\frac{1}{\tau\sqrt{\pi}}}\exp\left(-\frac{t^2}{2\tau^2}\right) $. The probability is computed for different values of the coupling constant $g$ where $ g_L = g_R \equiv g $ is assumed. The result of the computation is shown in Fig. \ref{fig:adiabatic}, which indicates that the flip probability is close to 1 if $ \kappa\tau \gtrsim 10 $.

\section{\label{sec:field}GENERATION OF ENTANGLEMENT BETWEEN ATOMS}

\begin{figure}
\centering \includegraphics[height=0.9\columnwidth, angle=270]{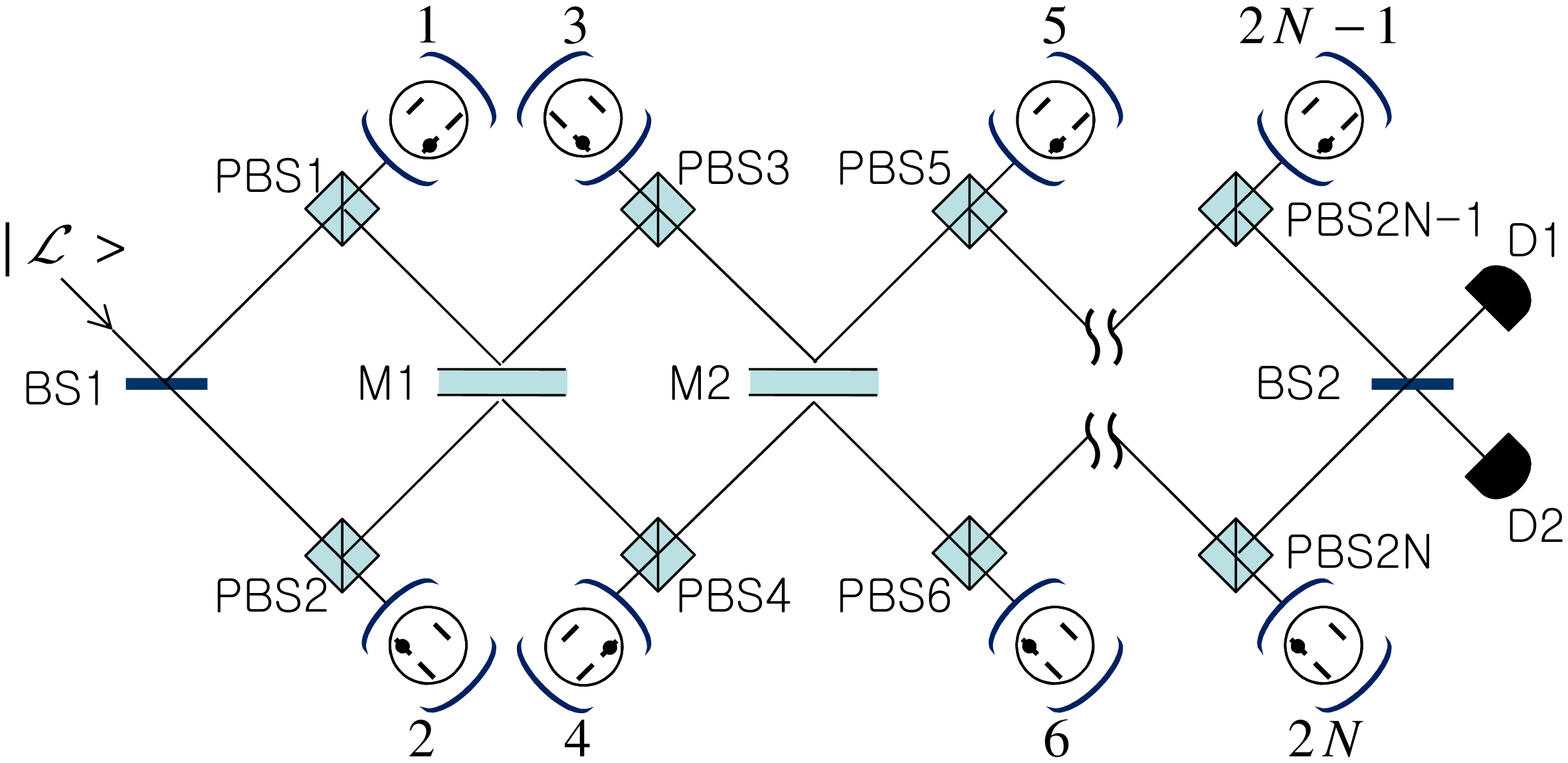}
\caption{\label{fig:GHZscheme} Scheme to generate 2N-atom GHZ state. BS, PBS, M and D refer respectively to a 50/50 beam splitter, polarizing beam splitter, mirror and detector. We assume that a phase shift of $\pi$ occurs when a photon is reflected from the lower surface of each beam splitter and that each PBS transmits a left-circularly polarized photon and reflects a right-circularly polarized photon. The lines in the figure that connect optical devices can be considered to be optical fibers. The far side of each cavity is assumed to be perfectly reflecting. The photon incident on beam splitter BS1 is left circularly polarized $\ket{{\cal L}}$ and atoms $1, 2, \cdots, 2N-1, 2N$ are prepared in state $\ket{LLRRLL\cdots}$.}
\end{figure}

We now present the actual arrangement of our scheme. Shown in Figure \ref{fig:GHZscheme} is the scheme that generates the GHZ state among 2N atoms each trapped in a cavity.
The atoms $1, 2, \cdots, 2N-1, 2N$ are prepared initially in state $\ket{LLRRLL\cdots}$ and a left-circularly polarized photon $\ket{{\cal L}}$ is directed to beam splitter BS1. Each cavity in the figure with a photon incident on it constitutes the basic building block depicted in Figure \ref{fig:atomblock}. Thus, if the photon is reflected (transmitted), then all odd(even)-numbered atoms in the upper(lower) row will experience a state flip $\ket{L}\leftrightarrow\ket{R}$. The presence of beam splitter BS2 ensures that which path information remains hidden. One can then easily deduce that detection of a photon at detector D1 (D2) signals that a 2N-atom entangled state $\frac{1}{\sqrt{2}}\left(\ket{RLLRRL\cdots}\pm\ket{LRRLLR\cdots}\right)$ is produced. Relabeling of the qubit states on each even- or odd-numbered atom, or a NOT operation on each even- or odd-numbered atom, then yields the standard 2N-atom GHZ state $\frac{1}{\sqrt{2}}\left(\ket{RRRRRR\cdots}\pm\ket{LLLLLL\cdots}\right)$.

\begin{figure}
\centering \includegraphics[height=0.9\columnwidth, angle=270]{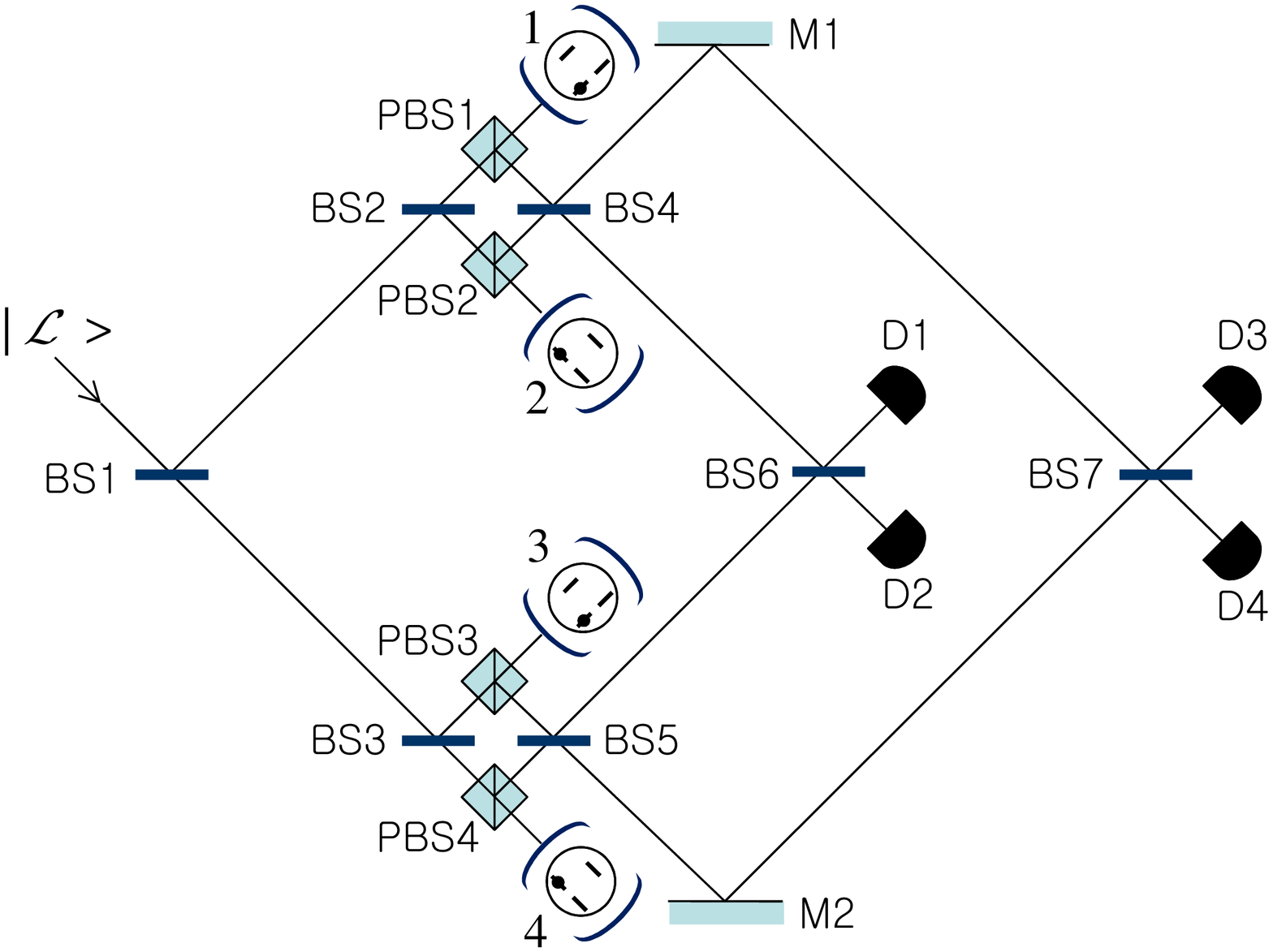}
\caption{\label{fig:4Wscheme} Scheme to generate four-atom {\it W} state. A left-circularly polarized photon is incident on beam splitter BS1 and atoms 1, 2, 3, 4 are prepared in state $\ket{LLLL}$.}
\end{figure}

\begin{figure}
\centering \includegraphics[height=0.9\columnwidth, angle=270]{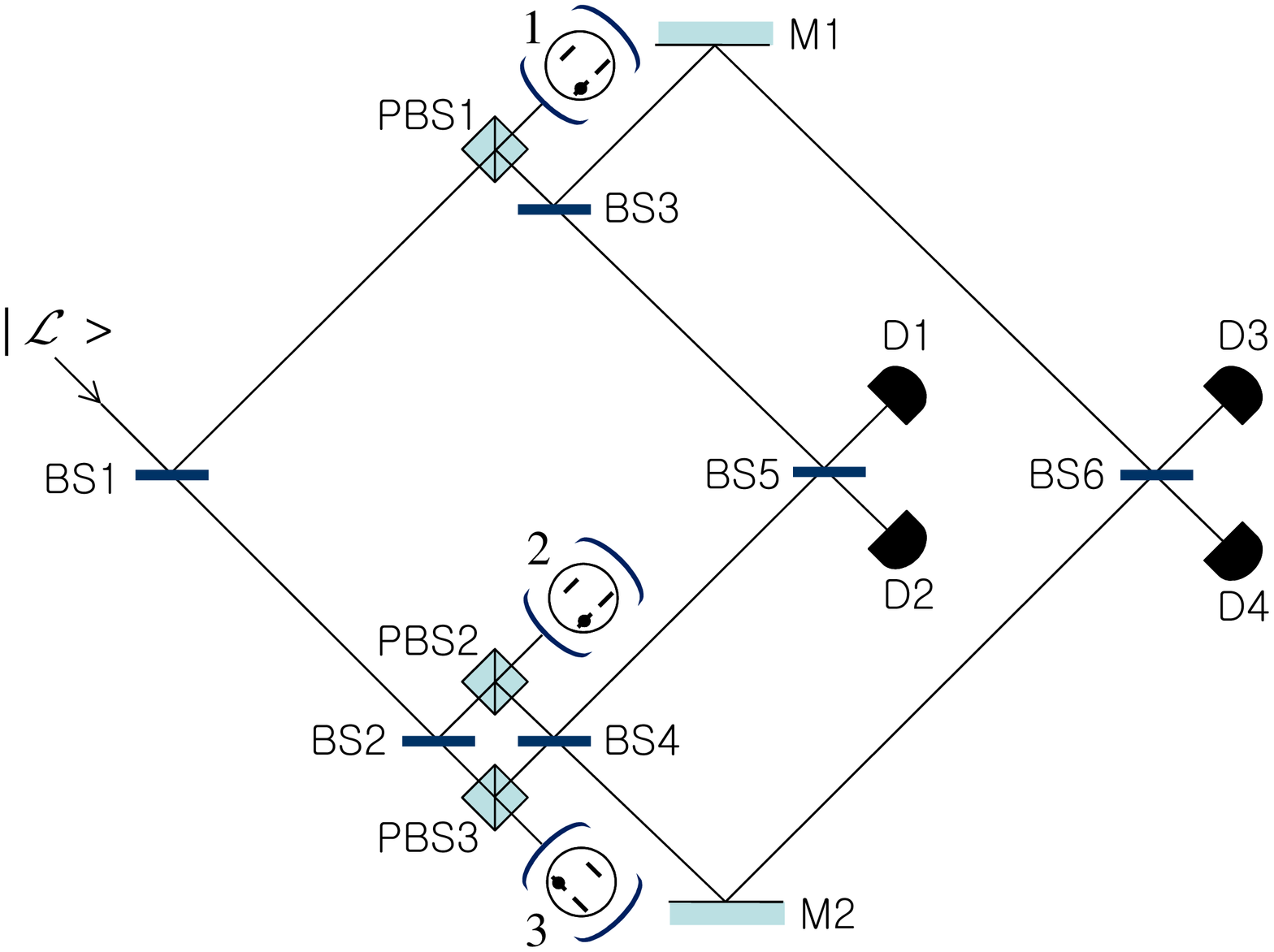}
\caption{\label{fig:3Wscheme} Scheme to generate three-atom {\it W} state. A left-circularly polarized photon is incident on beam splitter BS1 and atoms 1, 2, 3 are prepared in state $\ket{LLL}$. The reflectivity of BS1 is $\frac{1}{3}$ and that of other beam splitters is $\frac{1}{2}$.}
\end{figure}

Schemes to generate different types of entangled sates can easily be constructed with slight rearrangements of the system configuration. An example is given in Figure \ref{fig:4Wscheme} in which a scheme to generate a four-atom {\it W} state is shown.
The four atoms 1, 2, 3, 4 are each prepared initially in state $\ket{L}$ and a left-circularly polarized photon $\ket{{\cal L}}$ is directed to beam splitter BS1. It can be easily seen that detection of a photon at any of the four detectors D1, D2, D3 and D4 indicates that the atomic state flip $\ket{L}\rightarrow\ket{R}$ has occurred in one of the four atoms. Since it is unknown at which atom the flip has occurred, the state produced is of the form $\frac{1}{2}\left(\ket{RLLL}+\ket{LRLL}+\ket{LLRL}+\ket{LLLR}\right)$ within local operations. This scheme can easily be generalized to generate $N$-atom {\it W} state when $N=2^n$. If $N\not=2^n$, the $N$-atom {\it W} state can still be generated probabilistically. For example, by removing one of the cavities of Figure \ref{fig:4Wscheme} and the atom in that cavity and by replacing it with a detector D5, a three-atom {\it W} state can be generated with a success probability of $\frac{3}{4}$. The generation fails when D5 detects a photon, which occurs with a probability of $\frac{1}{4}$. If beam splitters of appropriate reflectivity's are available, deterministic generation can be achieved. For example, the scheme shown in Figure \ref{fig:3Wscheme} can generate the three-atom {\it W} state with 100\% probability, if the first beam splitter BS1 has a reflectivity of $\frac{1}{3}$ and all other beam splitters are the usual 50/50 beam splitters.

\begin{figure}
\centering \includegraphics[height=0.9\columnwidth, angle=270]{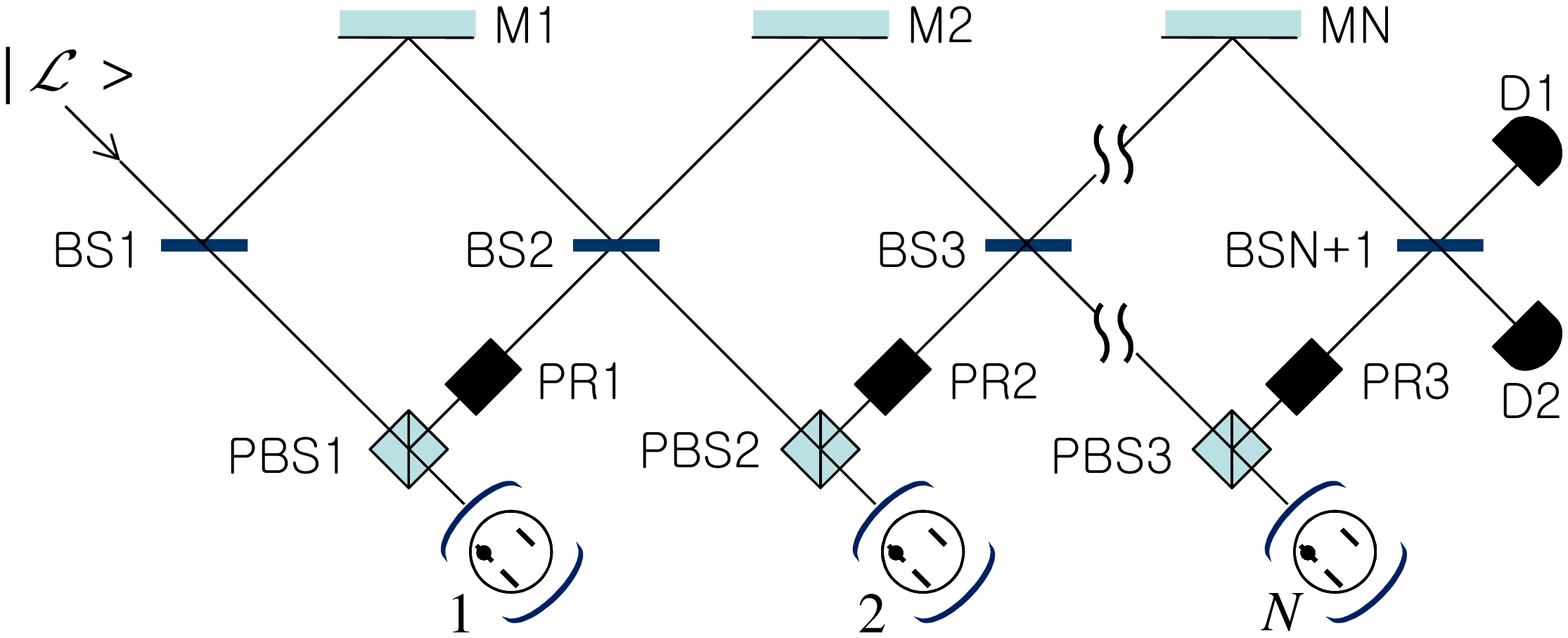}
\caption{\label{fig:clusterscheme} Scheme to generate linear cluster state. PR refers to polarization rotator that performs the NOT operation $\ket{{\cal L}}\leftrightarrow\ket{{\cal R}}$ on a photon. A left-circularly polarized photon is incident on beam splitter BS1 and atoms $1, 2, \cdots, N$ are prepared in $\ket{LL\cdots L}$.}
\end{figure}

In Figure \ref{fig:clusterscheme} we show an arrangement that generates $N$-atom linear cluster state.
As before all $N$ atoms are prepared each in $\ket{L}$ and a left-circularly polarized photon $\ket{{\cal L}}$ is incident on the beam splitter BS1. A new element in this arrangement is a polarization rotator that performs NOT operation $\ket{{\cal L}}\leftrightarrow\ket{{\cal R}}$ on the photon that enters it. Each polarization rotator ensures that the photon entering each cavity is left circularly polarized $\ket{{\cal L}}$. Noting that a phase shift of $\pi$ occurs when a photon is reflected from the lower side of each beam splitter, it is easy to verify that the atomic state generated upon detection of a photon at detector D1[D2] is given by $\left[Z_N\right]\prod_{i=1}^{N-1}CZ_{i,i+1}\left(\ket{L}_{i}+\ket{R}_{i}\right)\left(\ket{L}_{N}+\ket{R}_{N}\right)$, which is the $N$-atom linear cluster state. Here, $CZ_{i,i+1}$ represents a controlled-Z operation on the atomic pair $i$ and $i+1$. $Z_n$, $\sigma_Z$ (Pauli) operation on the last atom $N$, needs to be applied if the photon arrives at detector D2.

\section{GENERATION OF ENTANGLEMENT BETWEEN CAVITY FIELDS}

\begin{figure}
\centering \includegraphics[height=0.8\columnwidth, angle=270]{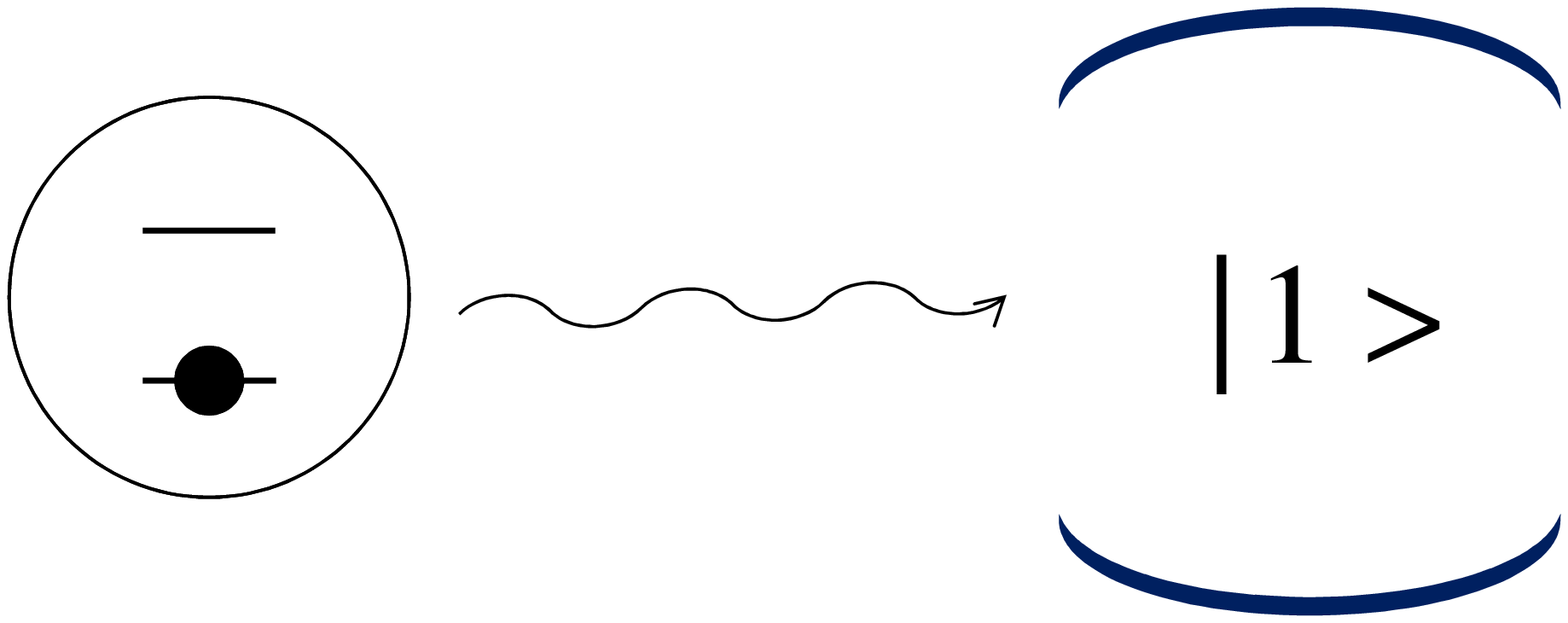}
\caption{\label{fig:fieldblock} The basic building block of the scheme to generate entanglement between cavity fields. The interaction between a cavity field prepared in a single-photon state $\ket{1}$ and a two-level atom in the ground state $\ket{g}$ transforms the system to the atom in the excited state $\ket{e}$ and the cavity field in vacuum $\ket{0}$, and vice versa.}
\end{figure}

Although generation of entangled atoms and photons has been investigated much in the past, relatively little attention has been given to generation of entanglement between cavity fields \cite{bib:prev6, bib:exp6}. One advantage of the scheme proposed in the previous section is that it can easily be modified to generate entanglement between cavity fields instead of between atoms. All that is needed essentially is a role switching between atoms and cavity fields. Now an atom, not a photon, gets to play the role of a flying qubit. As a basic building block, one needs a system that performs NOT operation on the cavity field state $\ket{0}\leftrightarrow\ket{1}$, where $\ket{0}$ and $\ket{1}$ are the vacuum and one-photon states. For the building block of our scheme, we choose an atom-cavity system of Figure \ref{fig:fieldblock}, where we assume that the atom and cavity parameters are chosen in such a way that the atom-field interaction corresponds to a $\pi$-pulse interaction.
The system then performs the state transformation $\ket{g}\ket{1}\leftrightarrow\ket{e}\ket{0}$, i.e., the field state flip $\ket{1}\leftrightarrow\ket{0}$ is achieved through the interaction with an atom. Note that here we choose a two-level atom as the flying qubit.

\begin{figure}
\centering \includegraphics[height=0.9\columnwidth, angle=270]{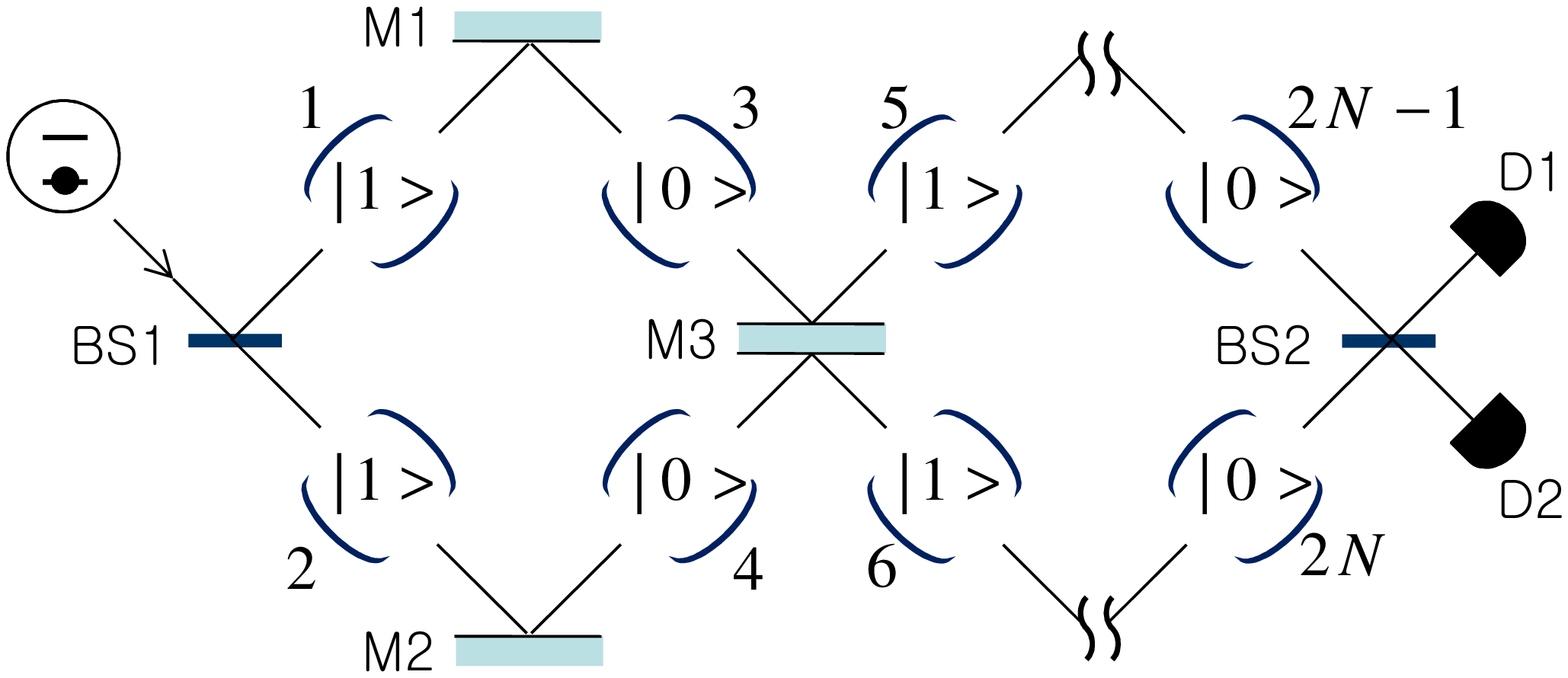}
\caption{\label{fig:fieldscheme} Scheme to generate 2N field GHZ state. BS, M, and D refer now to atomic beam splitter, atomic mirror and atomic detector. The atom incident on the atomic beam splitter BS1 is prepared in the ground state $\ket{g}$ and cavity fields $1, 2, \cdots, 2N-1, 2N$ are prepared in $\ket{110011\cdots}$.}
\end{figure}

Figure \ref{fig:fieldscheme} shows a scheme to generate the GHZ state among 2N cavity fields.
The cavity fields $1, 2, \cdots, 2N-1, 2N$ are prepared initially in the state $\ket{110011\cdots}$ and the atom, the flying qubit, in its lower state $\ket{g}$. The atomic beam splitter BS1 generates two possible paths for the atom and the second atomic beam splitter BS2 ensures that which path information remains hidden. With detection of an atom at the detector D1 (D2), the cavity field state assumes the 2N-field GHZ state $\frac{1}{\sqrt{2}}\left(\ket{011001\cdots}\pm\ket{100110\cdots}\right)$, which reduces to the standard GHZ state $\frac{1}{\sqrt{2}}\left(\ket{0000\cdots}\pm\ket{1111\cdots}\right)$ with local operations.

As was the case for the scheme of Figure \ref{fig:GHZscheme}, the scheme of Figure \ref{fig:fieldscheme} can also be modified easily to generate different types of entanglement, e.g., the {\it W} state and the cluster state, between cavity fields. For generation of the cluster state, we need, in addition to atomic beam splitters and atomic mirrors, a device that plays the role of the polarization rotator of Figure \ref{fig:clusterscheme}, i.e., a device that performs NOT operation $\ket{g}\leftrightarrow\ket{e}$ on the atomic qubit. This can be achieved by a cavity in a one-photon state $\ket{1}$ or in a vacuum state $\ket{0}$, depending on whether the incident atom is in $\ket{g}$ or $\ket{e}$, with parameters chosen to satisfy the $\pi$-pulse interaction time.

\section{GENERATION OF FIELD GRAPH STATES}

\begin{figure}
\centering \includegraphics[height=0.9\columnwidth, angle=270]{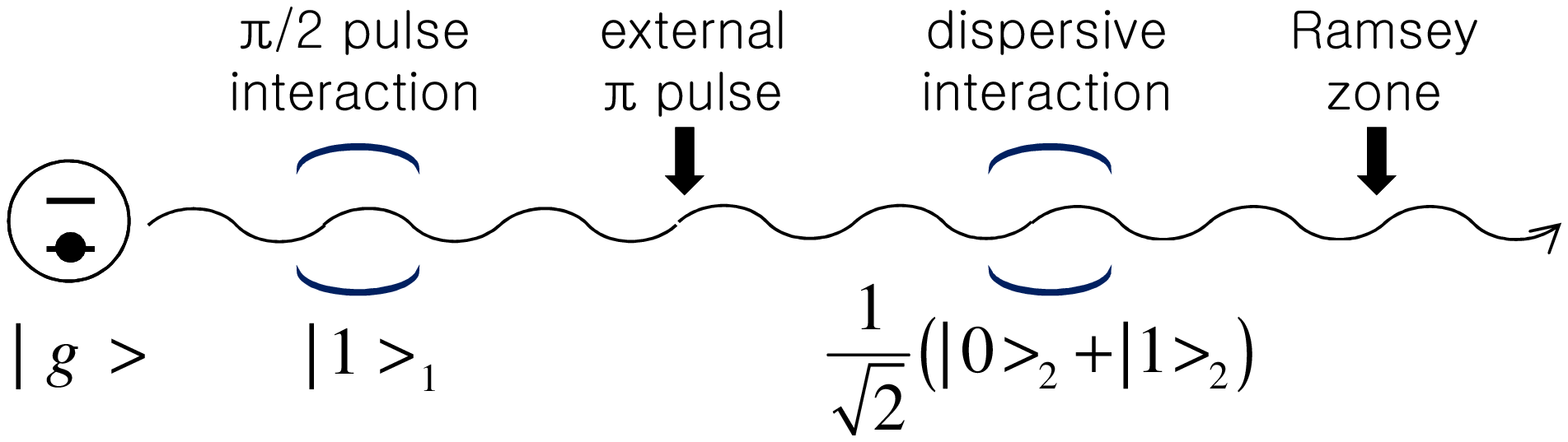}
\caption{\label{fig:graphscheme} Generation of the two-qubit cluster state between two cavity fields. A two-level atom in its lower level $\ket{g}$ goes through in turn, cavity 1 where $\frac{\pi}{2}$-pulse interaction occurs, an external $\pi$ pulse, cavity 2 where a dispersive interaction occurs, and a Ramsey zone.}
\end{figure}

In this section we introduce another scheme which is particularly suited to generate the field cluster states and, more generally, any type of the field graph state \cite{bib:graph}. The main idea stems from the observation that controlled-Z operation, which is a key operation required to generate graph states, between two cavity fields can be accomplished by entangling one cavity field with an atom and then by letting this atom go through a dispersive interaction with a second cavity field and pass through a Ramsey zone. This process is shown schematically in Figure \ref{fig:graphscheme}.
We first let a two-level atom prepared in its lower state $\ket{g}$ interact with the field of cavity 1 prepared in $\ket{1}_1$ through a $\frac{\pi}{2}$-pulse interaction. We then let the atom interact with an external $\pi$ pulse. The state of the atom and the cavity field 1 after the interaction is $\frac{1}{\sqrt{2}}\left(\ket{g}\ket{0}_1+\ket{e}\ket{1}_1\right)$. The atom is then allowed to pass through cavity 2, prepared in a symmetric superposition $\frac{1}{\sqrt{2}}\left(\ket{0}_2+\ket{1}_2\right)$, where state transformation $\ket{g}\ket{0}_2\rightarrow\ket{g}\ket{0}_2$, $\ket{g}\ket{1}_2\rightarrow\ket{g}\ket{1}_2$, $\ket{e}\ket{0}_2\rightarrow\ket{e}\ket{0}_2$, $\ket{e}\ket{1}_2\rightarrow-\ket{e}\ket{1}_2$ occurs through dispersive interaction. The state of the atom and the cavity fields 1, 2 then becomes $\frac{1}{2}\left(\ket{g}\ket{0}_1\ket{0}_2+\ket{g}\ket{0}_1\ket{1}_2+\ket{e}\ket{1}_1\ket{0}_2-\ket{e}\ket{1}_1\ket{1}_2\right)$. We now let the atom pass through a Ramsey zone where the state transformation $\ket{g}\rightarrow\frac{1}{\sqrt{2}}\left(\ket{g}+\ket{e}\right)$, $\ket{e}\rightarrow\frac{1}{\sqrt{2}}\left(\ket{g}-\ket{e}\right)$ is performed. The final state of the system then becomes
\setlength\arraycolsep{2pt}\begin{eqnarray*}
\frac{1}{2\sqrt{2}}[& \ket{g}\left(\ket{0}_1\ket{0}_2+\ket{0}_1\ket{1}_2+\ket{1}_1\ket{0}_2-\ket{1}_1\ket{1}_2\right)\\
+ & \ket{e}\left(\ket{0}_1\ket{0}_2+\ket{0}_1\ket{1}_2-\ket{1}_1\ket{0}_2+\ket{1}_1\ket{1}_2\right)].
\end{eqnarray*}
Detection of the atom in $\ket{g}$ or $\ket{e}$ ensures that the two cavity fields 1, 2 are in the two-qubit cluster state (which is locally equivalent to a Bell state).

With controlled-Z operation between cavity fields readily available as described above, one can easily construct a scheme to generate any desired field graph state. Consider, for example, the $N$-qubit star-type graph state which is equivalent to the $N$-qubit GHZ state. Such a field state can be generated by preparing cavity 1 in state $\frac{1}{\sqrt{2}}\left(\ket{0}_1+\ket{1}_1\right)$ and other $(N-1)$ cavities each in an entangled state with two-level atom, $\frac{1}{\sqrt{2}}\left(\ket{0_{j}g_{j}}+\ket{1_{j}e_{j}}\right)~(j=2, 3, \cdots, N-1, N)$, and then letting each atom one after another interact dispersively with the field of cavity 1 and pass through a Ramsey zone. Detection of $(N-1)$ atoms in any combination of states ensures generation of the $N$-qubit star-type field graph state apart from local unitary transformations. This therefore represents an alternative method to the scheme of Figure 7 to generate multipartite GHZ state among cavity fields. The advantage of the present method lies in the fact that it can easily be rearranged to generate other types of graph state. For example, the $N$-qubit linear graph state can be generated by preparing $N$ cavities in the same states as above and now letting atom $j(j=2,3,\cdots,N-1,N)$ one by one interact dispersively with the field of cavity $(j-1)$ and pass through a Ramsey zone. To take another example, the $N$-qubit ring-type graph state can be generated if one prepares all $N$ cavities in an entangled state $\frac{1}{\sqrt{2}}\left(\ket{0_{j}g_{j}}+\ket{1_{j}e_{j}}\right)~(j=2, 3, \cdots, N-1, N)$ and lets the atom $j(j=2, 3, \cdots, N-1, N)$ one by one interact dispersively with the field of cavity $(j-1)$ and pass through a Ramsey zone, and finally let the atom 1 interact dispersively with the field of cavity $N$ and pass through a Ramsey zone to close the ring.

\section{DISCUSSION}

The scheme we introduced in Sec. \ref{sec:field} for generation of atomic entanglement has several attractive features; it  requires only linear optical devices in addition to atom-cavity interaction and photon detection. It is scalable (it can be constructed to generate entanglement between an arbitrary number of atoms) and versatile (it covers a variety of multipartite entangled states such as the GHZ, {\it W} and cluster state; with a role switching between atoms and photons, the scheme can be modified to generate entanglement between cavity fields).

Another attractive feature of the scheme is that, although an efficient execution of the scheme requires adiabatic passage \cite{bib:adiabatic} for the state transformation $\ket{L}\ket{{\cal L}}\leftrightarrow\ket{R}\ket{{\cal R}}$ with 100\% success probability at each basic building block, it can still work even if the adiabatic condition is not met and the success probability is lower than 100\%. If the desired state flip does not occur and a ''wrong'' photon is emitted out of a cavity $j$, it moves backward to the previous cavity $(j-1)$ for another round of interaction. If a ''wrong'' photon is emitted out of cavity $(j-1)$ again, it moves backward to cavity $(j-2)$. On the other hand, if a ''right'' photon is emitted, it moves forward to cavity $j$ for another round of interaction. In other words, only the ''right'' photon can proceed forward, and thus detection of a photon at either detector D1 or D2 guarantees that the desired entangled state is produced, even though the photon may have moved back and forth many times on the way to the detector.

The need to use slow, adiabatic interactions in our scheme for generation of atomic entanglement means that the laser pulse, that plays the role of the flying qubit, must be sufficiently long. Fig. \ref{fig:adiabatic} suggests that the condition $ \kappa\tau \gtrsim 10 $ is to be met. As a consequence, the entire generation process may take an exceedingly long time, especially if the scheme involves a large number of cavities which is the case if the number of atoms to be entangled is large. Since the coherence of our atom-cavity system must be maintained throughout the generation process, the required slow interaction may be one important factor to limit the number of qubits that can be entangled by our scheme.

\begin{figure}
\centering \includegraphics[height=0.9\columnwidth, angle=270]{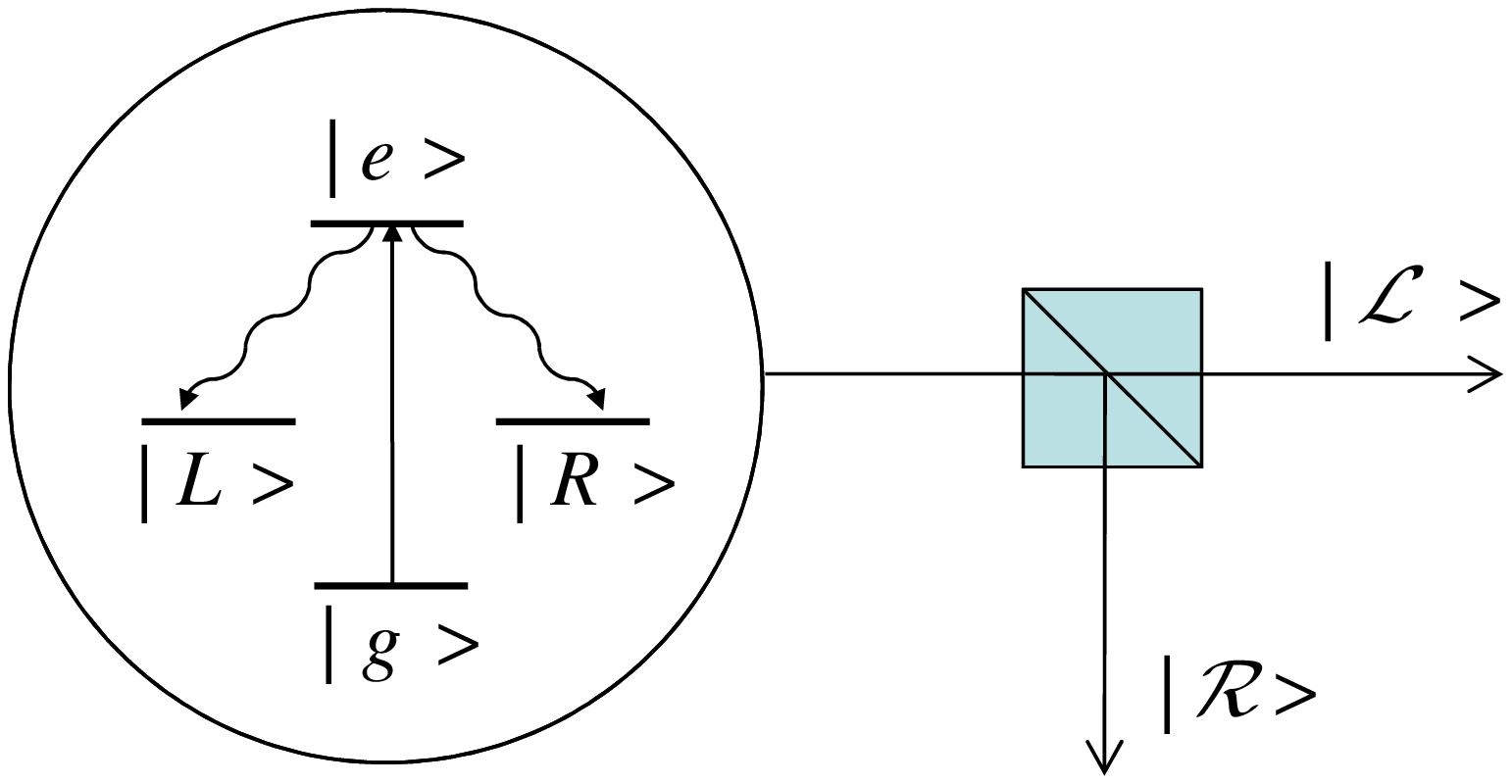}
\caption{\label{fig:injection} Single polarized-photon injection by using a fourth level.}
\end{figure}

The technological demand of our proposed scheme is high as it operates at a single photon level. The optical fibers used in the scheme should have sufficiently low loss, so that we have a high probability for the photon, flying qubit, to reach the detector. The demand on low loss is particularly high if entanglement is to be generated between a large number of atoms as the flying qubit needs to travel a long distance, or if the adiabatic condition is not met and the photon may have to move back and forth between the cavities many times. Nevertheless, it is one of the merits of the proposed scheme that the photon loss leads simply to no detection of a photon at the detectors and thus lowers the success probability of the entire generation scheme, but does not lead to an error or any wrong message. The requirement of initial single polarized-photon injection is also difficult to meet, as a true single photon source does not exist. One possible way of obtaining single-photon injection is to use an additional fourth level of our three-level atom, as shown in Figure \ref{fig:injection}.
The idea is to prepare the atom on the fourth level $\ket{g}$ and adiabatically drive it with classical field to the state $\ket{e}$ \cite{bib:prev2}. The atom will then decay spontaneously to $\ket{L}$ or $\ket{R}$ emitting a left- or right-circularly polarized photon. By using a polarizing beam splitter that transmits only the left-circularly polarized photon, we can obtain single polarized-photon injection probabilistically.

Our scheme requires an atom-cavity system to be constructed and operated in exactly the way to give the desired performance. It is particularly important to have precise control of atom-cavity interactions, which in turn requires easy and precise manipulation of the positions of single atoms in a cavity relative to the cavity mode, as well as easy loading of single atoms to cavities on demand. A high technical challenge in our scheme is encountered especially when it involves a large number of cavities, i.e., when a large number of qubits are to be entangled. In this case a cavity array needs to be constructed, and one may consider employing an optical microcavity network with adjacent cavities connected with optical fibers \cite{bib:cavity}. It is a difficult task to fabricate small cavities in large numbers, which offer easy and precise control of atom-cavity coupling. The construction of a cavity array requires also an ability to couple the cavities to optical fibers with high coupling efficiencies. Despite impressive progress made in recent years in cavity technology \cite{bib:trap}, it is still a high challenge to construct an array of atom-cavity systems that possess all these desired properties.

Our scheme of Figure 7 to generate field entangled states does not have high technical demand for initial injection as it requires a single atom as the flying qubit. On the other hand, the scheme requires a reliable operation of atomic beam splitters and atomic mirrors, which is more difficult to achieve than the optical counterpart \cite{bib:atomicbs1}. In addition, as the atom travels much more slowly than photons, the scheme may be difficult to implement if the cavities to be entangled are separated far from one another.

As far as the atomic beam splitting and the atomic mirror are concerned, they have been demonstrated both theoretically \cite{bib:atomicbs2} and experimentally \cite{bib:atomicbs3}. In such kind of actions atomic Bragg diffraction is the appropriate tool. By sending the atoms with Bragg angle and choosing the interaction times of atom with the cavity, one can efficiently control the beam splitting and the mirror action. For example, an atom with initial momentum state $\ket{p_0}(\ket{p_{-2}})$ under first order Bragg diffraction can have the state transformation $\ket{p_0}\rightarrow\left(\ket{p_0}+\ket{p_{-2}}\right)/\sqrt{2}$ [and similarly $\ket{p_{-2}}\rightarrow\left(\ket{p_0}-\ket{p_{-2}}\right)/\sqrt{2}$]. This action is the analogous part of the $\pi$ pulse in atomic internal states. And also the mirror action for a $\pi/2$ pulse can invert the momentum state as $\ket{p_0}\rightarrow\ket{p_{-2}}$ and $\ket{p_{-2}}\rightarrow\ket{p_0}$.\\

\begin{acknowledgements}
This research was supported by a Grant from Korea Research Institute for Standards and Science (KRISS). The authors thank Professor Suhail Zubairy of Texas A\&M University, U.S.A., and Dr. Jaeyoon Cho of Korea Research Institute for Standards and Science for useful discussions.
\end{acknowledgements}

\end{document}